\documentclass[twocolumn,prx,showpacs,preprintnumbers,amsmath,amssymb,citeautoscript,superscriptaddress]{revtex4-1}
\usepackage[latin9]{inputenc}
\usepackage{color}
\usepackage{amsmath}
\usepackage{amssymb}
\usepackage{graphicx}

\makeatletter
\usepackage{amsfonts}\usepackage{here}
\usepackage{dcolumn}
\usepackage{bm}
\usepackage{color}
\usepackage{here}

\newlength{\figwidth}
\newcommand{\BaRb}{Ba$_{1-x}$Rb$_{x}$Fe$_{2}$As$_{2}$}
\newcommand{\BaK}{Ba$_{1-x}$K$_{x}$Fe$_{2}$As$_{2}$}

\newcommand{\RbAs}{RbFe$_{2}$As$_{2}$}
\newcommand{\CsAs}{CsFe$_{2}$As$_{2}$}

\newcommand{\tc}{$T_{{\rm c}}$}

\makeatother

\begin{document}

\title{Emergent XY electronic nematicity in iron-based superconductors}

\author{K. Ishida}

\thanks{These authors contributed equally to this work.}

\author{M. Tsujii}

\thanks{These authors contributed equally to this work.}

\author{S. Hosoi}

\thanks{Present address: Department of Materials Engineering Science, Osaka
University, Toyonaka, Osaka 560-8531, Japan.}

\author{Y. Mizukami}

\affiliation{Department of Advanced Materials Science, University of Tokyo, Kashiwa,
Chiba 277-8561, Japan}

\author{S. Ishida}

\author{A. Iyo}

\author{H. Eisaki}

\affiliation{Electronics and Photonics Research Institute, National Institute
of Advanced Industrial Science and Technology, Tsukuba, Ibaraki 305-8568,
Japan}

\author{T. Wolf}

\author{K. Grube}

\author{H. v. L\"{o}hneysen}

\affiliation{Institut f\"{u}r Festk\"{o}rperphysik, Karlsruher Institut f\"{u}r Technologie,
76021 Karlsruhe, Germany}

\author{R. M. Fernandes}

\affiliation{School of Physics and Astronomy, University of Minnesota, Minneapolis,
Minnesota 55455, U.S.A.}

\author{T. Shibauchi}

\affiliation{Department of Advanced Materials Science, University of Tokyo, Kashiwa,
Chiba 277-8561, Japan}
\begin{abstract}
Electronic nematicity, a correlated state that spontaneously breaks
rotational symmetry, is observed in several layered quantum materials.
In contrast to their liquid-crystal counterparts, the nematic director
cannot usually point in an arbitrary direction (XY nematics), but
is locked by the crystal to discrete directions (Ising nematics),
resulting in strongly anisotropic fluctuations above the transition.
Here, we report on the observation of nearly isotropic XY-nematic
fluctuations, via elastoresistance measurements, in hole-doped Ba$_{1-x}$Rb$_{x}$Fe$_{2}$As$_{2}$
iron-based superconductors. While for $x=0$ the nematic director
points along the in-plane diagonals of the tetragonal lattice, for
$x=1$ it points along the horizontal and vertical axes. Remarkably,
for intermediate doping, the susceptibilities of these two symmetry-irreducible
nematic channels display comparable Curie-Weiss behavior, thus revealing
a nearly XY-nematic state. This opens a new route to assess this elusive
electronic quantum liquid-crystalline state, which is a candidate
to host unique phenomena not present in the Ising-nematic case.
\end{abstract}
\maketitle

\section{INTRODUCTION}

Liquid crystals are composed of molecules with anisotropic shapes,
which possess a degree of orientational order. As a result, they can
form nematic phases, in which translational invariance is preserved,
but rotational symmetry is spontaneously broken by the selection of
a director along which the molecules align. Strongly correlated electron
systems often show exotic states of matter analogous to liquid-crystalline
states, originating from nontrivial quantum many-body interactions.
In particular, electronic nematic phases have been found in a number
of quantum materials and have generated a lot of attention \cite{Fradkin2010,Fernandes2014}.
After its original experimental identification in quantum Hall systems
\cite{Lilly1999}, electronic nematicity has increasingly been recognized
as a ubiquitous feature of unconventional superconductors. In iron-based
superconductors, electronic nematic order and fluctuations span a
wide region of the phase diagram \cite{Chu2010,Kasahara2012}, and
likely impact superconductivity \cite{Lederer2016}. Electronically-driven
rotational symmetry breaking has also been reported inside the mysterious
pseudogap phase of high-$T_{{\rm c}}$ cuprates \cite{Hinkov2008,Daou2010,Sato2017}
as well as in the elusive hidden-order phase of the heavy-fermion
material URu$_{2}$Si$_{2}$ \cite{Okazaki2011}.

A crucial difference between electronic and liquid-crystalline nematic
orders is that the former develops in the presence of a crystal, which
by itself does break the rotational symmetry of the system \textendash{}
for instance, by restricting the possible paths along which the electrons
can hop. As a result, while in liquid crystals the nematic director
can point anywhere in space (i.e., it has a continuous symmetry),
in electronic systems it is generally restricted to a few high-symmetry
directions of the crystal (i.e., it has a discrete symmetry). While
such a discrete electronic nematic state still displays interesting
properties \cite{Fradkin2010,Fernandes2014}, a hypothetical continuous
electronic nematic state would host several unique quantum many-body
phenomena \cite{Oganesyan2001}. For instance, the transverse and
longitudinal fluctuations in the disordered phase have completely
different dynamics, which may lead to multiscale critical behavior
\cite{Zacharias2009}. Moreover, the ordered phase has a Goldstone
mode associated with the breaking of the continuous symmetry by the
director. Remarkably, unlike other Goldstone modes such as phonons
or magnons, the nematic mode couples directly to the electronic density
rather than to gradients of the density \cite{Watanabe2014}. This
can lead to strange metallic behavior \cite{Oganesyan2001} and strongly
impact superconductivity \cite{Kim2004}.

At first sight, the inevitable presence of the crystal may be taken
as an insurmountable obstacle to realize a continuous electronic nematic
state in actual materials. However, as we argue now, and demonstrate
below experimentally in the hole-doped iron pnictide Ba$_{1-x}$Rb$_{x}$Fe$_{2}$As$_{2}$,
a nearly-continuous nematic phase can be observed under certain conditions.
To make the discussion concrete, we consider layered crystals with
tetragonal symmetry, i.e., the symmetry of a square. Indeed, many
of the materials where nematicity has been observed do possess this
symmetry. We start our discussion by first assuming that there was
no underlying crystal. Then the electronic nematic director could
point anywhere in the layer plane. In this case, in analogy to a two-dimensional
nematic liquid crystal, the nematic order parameter is described in
terms of a traceless symmetric $2\times2$ matrix \cite{Oganesyan2001}
given by: $\boldsymbol{\mathcal{Q}}=\left(\begin{array}{rr}
N_{1} & N_{2}\\
N_{2} & -N_{1}
\end{array}\right)$. Here, the order parameter $N_{1}$ corresponds to a quadrupolar
charge density that makes the horizontal ($x$) and vertical ($y$)
directions inequivalent ($d_{x^{2}-y^{2}}$-wave form factor), whereas
$N_{2}$ corresponds to a quadrupolar charge density that makes the
two diagonal directions inequivalent ($d_{xy}$-wave form factor).
A linear combination of $N_{1}$ and $N_{2}$ can be constructed such
that the quadrupolar charge density can be rotated along any in-plane
direction. The static and uniform free energy depends only on traces
of even powers of $\boldsymbol{\mathcal{Q}}$, i.e., only on the combination
$\left(N_{1}^{2}+N_{2}^{2}\right)$, reflecting the continuous symmetry
of the nematic director. Therefore, $N_{1}$ and $N_{2}$ can be interpreted
as two orthogonal components of the continuous nematic order parameter.
Borrowing the nomenclature of magnetic phases, the resulting nematic
phase is called an XY-nematic state.

The existence in reality of the tetragonal crystal, of course, dramatically
changes this scenario. Now, $N_{1}$ and $N_{2}$ transform as two
different irreducible representations of the tetragonal point group,
denoted respectively by $B_{1g}$ and $B_{2g}$. Phenomenologically,
this leads to a term in the free energy that explicitly lifts the
degeneracy between $N_{1}$ and $N_{2}$, changing the symmetry of
the nematic state from XY (continuous) to Ising (discrete). More formally,
the free energy expansion becomes

\begin{equation}
F=\frac{\gamma}{2}(N_{1}^{2}-N_{2}^{2})+\frac{a}{2}(N_{1}^{2}+N_{2}^{2})+\frac{u}{4}(N_{1}^{2}+N_{2}^{2})^{2}.\label{eq1}
\end{equation}

Here, $a$ and $u$ are Landau coefficients that depend on the microscopic
model. The anisotropic term $\gamma$ selects one of the two components
of the XY nematic order parameter: $B_{1g}$ nematic order when $\gamma<0$,
and $B_{2g}$ nematic order when $\gamma>0$. The situation is analogous
to an XY (i.e., planar) ferromagnet with an easy-axis anisotropy.
Indeed, the nematic order parameter can now be written in terms of
Pauli matrices as $\boldsymbol{\mathcal{Q}}=N_{1}\boldsymbol{\sigma}_{z}+N_{2}\boldsymbol{\sigma}_{x}$,
expressing the analogy between $N_{i}$ and magnetic moment components.
In the magnetic analogue, the $\gamma$ term corresponds to the single-ion
anisotropy that defines the magnetization easy axis. This analogy
offers a fruitful insight: in ferromagnets, when the single-ion anisotropy
is much smaller than the exchange coupling, although the magnetic
transition is formally Ising-like, the fluctuation spectrum is very
similar to that of the XY ferromagnet, except at very low energies.
Moreover, once long-range order sets in, the collective modes correspond
essentially to a Goldstone mode with a very small gap.

Thus, going back to the nematic case, it is expected that many of
the properties of the XY-nematic state will be relevant to describe
the system's behavior over a wide energy range as long as the anisotropy
coefficient $\left|\gamma\right|$ is much smaller than the nematic
energy scale, which we can set to be the nematic transition temperature
$T_{{\rm nem}}$. The remaining task is to find a concrete system
with $|\gamma|\ll T_{{\rm nem}}$. A promising strategy is to interpolate
between two different compounds that display $B_{1g}$ nematic order
($\gamma<0$) and $B_{2g}$ nematic order ($\gamma>0$), forcing $\gamma$
to cross zero. In the remainder of this paper, we show that such a
situation is realized in the hole-doped iron pnictide Ba$_{1-x}$Rb$_{x}$Fe$_{2}$As$_{2}$.

\section{METHODS}

The parent compound BaFe$_{2}$As$_{2}$ ($x=0$) displays an antiferromagnetic
transition that is preceded by a tetragonal to orthorhombic structural
transition, which makes the diagonals $[110]$ and $[1\bar{1}0]$
of the tetragonal crystallographic unit cell inequivalent. It is well
established that the structural transition is not driven by lattice
degrees of freedom \cite{Chu2012}, but is the result of an electronic
nematic instability in the $B_{2g}$ channel along the Fe-Fe direction
{[}Fig.\,1(a){]}. Note that our notation refers to the crystallographic
2-Fe unit cell; in the 1-Fe unit cell notation sometimes used in the
literature, $B_{1g}$ and $B_{2g}$ are exchanged. Upon increasing
doping, the structural transition temperature is suppressed, approaching
a putative quantum critical point near optimal doping, where the superconducting
dome is peaked. The superconducting dome extends all the way up to
$x=1$. Interestingly, in all hole-doped systems Ba$_{1-x}$$A_{x}$Fe$_{2}$As$_{2}$,
with alkali metal $A=$ K, Rb, Cs, the effective electronic mass is
observed to be strongly enhanced near $x=1$ \cite{Hardy13,Eilers2016,Hardy2016,Mizukami2016}.
This suggests a progressively more important role of electronic correlations
in the $3d^{5.5}$ electronic configuration ($x=1$) as compared to
the $3d^{6}$ configuration ($x=0$), which may be related to proximity
to a hypothetical $3d^{5}$ Mott insulating state \cite{Medici2014,Valenti15}.
This motivates a careful investigation of nematic tendencies in extremely
hole-doped pnictides by using high-quality single crystals (see Appendix
A), with the prospect of finding a suitable alloy series with a crossover
to $B_{1g}$ dominance, as the stronger correlations may promote a
mechanism for electronic nematicity different from that of the undoped
compound \cite{Kivelson1998}.

\begin{figure}[t]
\begin{centering}
\includegraphics[width=1\linewidth]{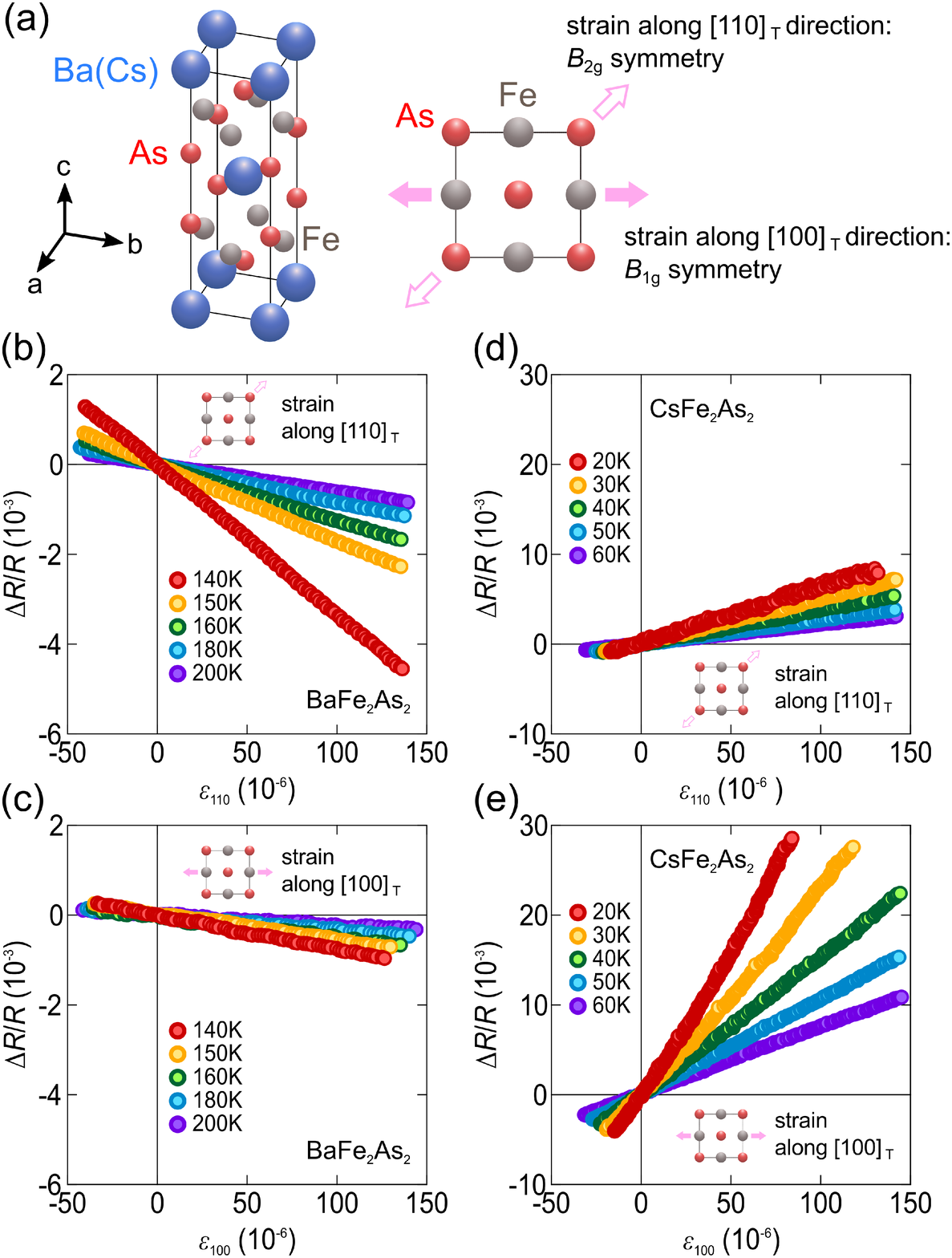} 
\par\end{centering}
 \caption{Elastoresistance under strain for two symmetry channels in $A$Fe$_{2}$As$_{2}$
($A=$ Ba and Cs). (a) Crystal structure of the tetragonal $A$Fe$_{2}$As$_{2}$
($A=$ Ba, Cs, or Ba$_{1-x}$Rb$_{x}$). By applying the strain parallel
to the tetragonal {[}110{]} or {[}100{]} direction (nearest Fe-Fe
bond or Fe-As bond direction, respectively), the nematic susceptibilities
in the $B_{2g}$ or $B_{1g}$ channels can be extracted. (b),(c) Representative
elastoresistance data $\Delta R/R$ of BaFe$_{2}$As$_{2}$ samples
at several temperatures above $T_{{\rm nem}}$ as a function of strain
along {[}110{]} (b) and {[}100{]} (c) directions. The samples are
glued on piezo stacks and the strain is induced by applying voltage
to the piezo stacks. The current direction is parallel to the strain
direction in both configurations. (d),(e) Similar data for CsFe$_{2}$As$_{2}$
for the {[}110{]} (d) and {[}100{]} (e) orientations. }
\label{fig1} 
\end{figure}


To study nematicity in the extremely hole-doped regime, we use elastoresistance
measurements, which have been shown to probe the bare static nematic
susceptibility \cite{Chu2012,Kuo2016,Hosoi2016}. In this technique,
we measure the relative change in resistance $\Delta R/R$ in response
to an externally applied strain $\epsilon$, which is controlled by
a piezoelectric device. Because uniaxial strain couples linearly to
the nematic order parameters, in the linear response regime the quantity
$\chi_{{\rm nem}}\sim\frac{1}{R}\frac{\partial\Delta R}{\partial\epsilon}$
is proportional to the nematic susceptibility (see Appendix B). Specifically,
in the iron pnictides, the nematic susceptibilities in the $B_{1g}$
and $B_{2g}$ symmetry channels (denoted here by $\chi_{{\rm nem}}^{[100]}$
and $\chi_{{\rm nem}}^{[110]}$, respectively, with the superscript
denoting the strain direction) can be obtained by applying strain
along the nearest neighbor Fe-As and Fe-Fe bond directions, corresponding
to $[100]$ and $[110]$, respectively {[}Fig.\,1(a){]}.

\section{RESULTS AND DISCUSSION}

Figure\,1(b)-(e) shows representative data of $\Delta R/R$ as a
function of applied strain $\epsilon$ along the $[100]$ and $[110]$
directions. In BaFe$_{2}$As$_{2}$, which displays the $3d^{6}$
electronic configuration, the elastoresistances in the two channels
show a striking anisotropy, with the response to the strain along
the $[110]$ direction being much larger than that along $[100]$
{[}Fig.\,1(b),(c){]}. This is consistent with previous reports \cite{Chu2012},
and reflects the Ising-like character of the $B_{2g}$ nematic state,
which sets in when $\chi_{{\rm nem}}^{[110]}$ peaks. In sharp contrast,
in CsFe$_{2}$As$_{2}$ with $3d^{5.5}$ electronic configuration,
the elastoresistance for strain applied along $[100]$ becomes much
larger than that along $[110]$, indicating that Ising-like $B_{1g}$
nematic fluctuations are dominant. Such a difference can be clearly
seen in the temperature dependence of the nematic susceptibilities
obtained from the slopes of $\Delta R/R$ as function of $\epsilon$,
as shown in Fig.\,2(a),(b). Although the nematic susceptibilities
presented here were evaluated from longitudinal elastoresistance measurements,
we confirmed that they are identical to those determined by using
a modified Montgomery method, indicating negligible contaminations
of $A_{1g}$-symmetry strain (see Appendix:B).

\begin{figure}[t]
\begin{centering}
\includegraphics[width=1\linewidth]{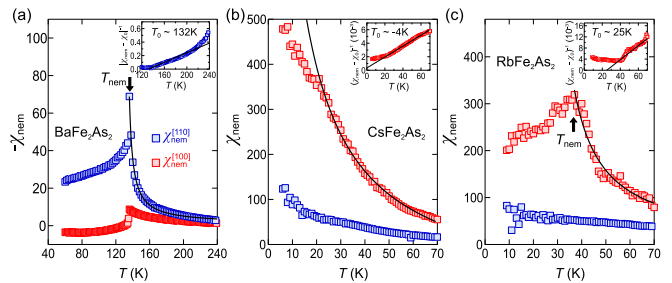} 
\par\end{centering}
\caption{Temperature dependence of the nematic susceptibility $\chi_{{\rm nem}}^{[100]}$
($B_{1g}$ or Fe-As direction, red squares) and $\chi_{{\rm nem}}^{[110]}$
($B_{2g}$ or Fe-Fe direction, blue squares) in BaFe$_{2}$As$_{2}$
(a), CsFe$_{2}$As$_{2}$ (b), and RbFe$_{2}$As$_{2}$ (c), as determined
by the elastoresistance measurements along the {[}100{]} and {[}110{]}
directions. The solid lines are the Curie-Weiss fits for the dominant
nematic fluctuations. The insets show the temperature dependence of
the inverse susceptibility, whose intercept provides an estimate of
the Curie-Weiss temperature $T_{0}$. }
\label{fig2} 
\end{figure}


In the case of BaFe$_{2}$As$_{2}$ {[}Fig.\,2(a){]}, $\chi_{{\rm nem}}^{[110]}$
is much larger and has a more pronounced temperature dependence as
compared with $\chi_{{\rm nem}}^{[100]}$. Moreover, as demonstrated
previously \cite{Chu2012}, $\chi_{{\rm nem}}^{[110]}$ has a peak
at the nematic transition temperature $T_{\mathrm{nem}}$. Above this
temperature, the nematic susceptibility can be described by a Curie-Weiss
expression 
\begin{equation}
\chi_{{\rm nem}}=\chi_{0}+\frac{C}{T-T_{0}}.\label{eq2}
\end{equation}
Here, $C$ is a constant, and $\chi_{0}$ is a temperature-independent
term not related to the nematic fluctuations. The Curie-Weiss temperature
$T_{0}$ represents the ``bare'' nematic transition temperature
without nemato-lattice coupling effects. Such a $B_{2g}$ Ising-nematic
order has been observed in many iron-based superconductors \cite{Kuo2016,Hosoi2016}.
However, in CsFe$_{2}$As$_{2}$ {[}Fig.\,2(b){]}, we find that the
dominant nematic fluctuations appear in the $B_{1g}$ channel, since
$\chi_{{\rm nem}}^{[100]}$ displays a considerably larger magnitude
and stronger temperature dependence than $\chi_{{\rm nem}}^{[110]}$.
We also note that the sign of $\chi_{{\rm nem}}^{[100]}$ is positive
for CsFe$_{2}$As$_{2}$, in contrast to the negative sign of $\chi_{{\rm nem}}^{[110]}$
in BaFe$_{2}$As$_{2}$. Such a sign reversal has also been reported
for the in-plane resistivity anisotropy in Ba$_{1-x}$K$_{x}$Fe$_{2}$As$_{2}$,
where the hole doping changes sign of $\rho_{b}/\rho_{a}-1$ from
positive to negative \cite{Blomberg2013}.

As shown in Fig.\,2(c), similar $B_{1g}$ Ising-nematic behavior
as in CsFe$_{2}$As$_{2}$ is also observed in RbFe$_{2}$As$_{2}$,
suggesting that nematicity along the Fe-As direction, 45$^{\circ}$
tilted with respect to the Fe-Fe direction, is a generic feature of
the $3d^{5.5}$ electronic configuration. Indeed, recent nuclear magnetic
resonance and scanning tunneling microscopy studies suggested $B_{1g}$
nematicity in these compounds \cite{Li2016,Liu2018}, although the
presence of long-range nematic order has not been settled. Here, we
note that not only do the $\chi_{{\rm nem}}^{[100]}(T)$ curves for
both CsFe$_{2}$As$_{2}$ and RbFe$_{2}$As$_{2}$ follow Curie-Weiss
behavior over a wide temperature range, but also does the nematic
susceptibility of RbFe$_{2}$As$_{2}$ exhibit a distinct peak at
$T_{{\rm nem}}\sim38$\,K, indicative of the onset of long-range
nematic order in the $B_{1g}$ channel. Recent specific-heat measurements
under in-plane field rotation indeed show two-fold oscillations near
$T_{c}$, supporting the $B_{1g}$ nematic order in RbFe$_{2}$As$_{2}$
\cite{Tanaka}. This is consistent with a positive Curie-Weiss temperature
$T_{0}\approx25$\,K obtained for RbFe$_{2}$As$_{2}$ by fitting
Eq.\,(\ref{eq2}) to $\chi_{{\rm nem}}^{[100]}(T)$. In contrast,
we have found $T_{0}\approx-4$\,K for CsFe$_{2}$As$_{2}$, from
which we conclude that CsFe$_{2}$As$_{2}$ is close to a possible
$B_{1g}$ nematic quantum critical point (QCP), which would correspond
to $T_{0}=0$. The largely enhanced magnitude of $\chi_{{\rm nem}}^{[100]}$
is also consistent with quantum-critical fluctuations of $B_{1g}$
nematicity \cite{Hosoi2016}. We note in passing that, at low temperatures,
$\chi_{{\rm nem}}^{[100]}$ in CsFe$_{2}$As$_{2}$ shows a noticeable
deviation from the Curie-Weiss behavior. Similar deviations have been
seen in several iron-based materials in the vicinity of their putative
$B_{2g}$ nematic QCP, when the Curie-Weiss temperature extrapolates
to zero \cite{Kuo2016,Hosoi2016}. Although disorder effects have
been discussed as a possible origin of this behavior, the observation
of a similar deviation in the very clean CsFe$_{2}$As$_{2}$ compound,
where sharp quantum oscillations are clearly observed \cite{Eilers2016},
suggests that mechanisms other than disorder may be important for
the behavior close to a nematic QCP.

Our elastoresistance measurements thus establish two parent iron-pnictide
compounds, the $3d^{6}$ BaFe$_{2}$As$_{2}$ and the $3d^{5.5}$
RbFe$_{2}$As$_{2}$, that display Ising-nematic orders in two orthogonal
channels, namely, $B_{2g}$ (corresponding to $\gamma>0$ in Eq.\,(\ref{eq1}))
and $B_{1g}$ (corresponding to $\gamma<0$ in Eq.\,(\ref{eq1})),
respectively. Following the strategy outlined above, we thus search
for signatures of XY nematicity in the doped compound Ba$_{1-x}$Rb$_{x}$Fe$_{2}$As$_{2}$.
Note that just because Ba$_{1-x}$Rb$_{x}$Fe$_{2}$As$_{2}$
interpolates between $B_{1g}$ and $B_{2g}$ nematic orders is not
enough to ensure XY nematicity. On the contrary, if these ordered
states had completely different origins, one would expect quite generically
the $B_{1g}$ fluctuations to be suppressed before $B_{2g}$ fluctuations
become sizable (and vice-versa). However, if these two states were
manifestations of a common nematic state, differing from each other
only in the direction of the nematic director, one would expect XY
nematicity to emerge in the regime where the director switches from
one direction to the other.

\begin{figure}[t]
\begin{centering}
\includegraphics[width=1\linewidth]{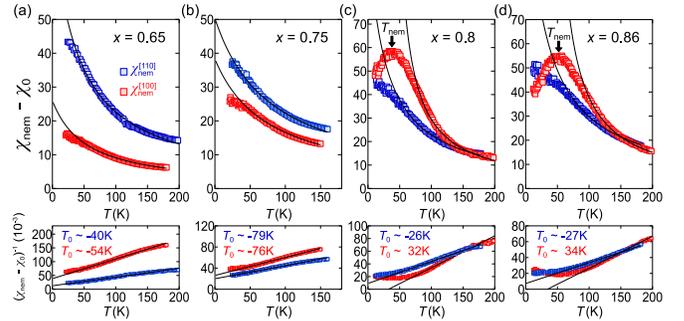} 
\par\end{centering}
\caption{Nematic susceptibilities in the two symmetry channels $\chi_{{\rm nem}}^{[100]}$
(red squares) and $\chi_{{\rm nem}}^{[110]}$ (blue squares) for Ba$_{1-x}$Rb$_{x}$Fe$_{2}$As$_{2}$
single crystals with $x=0.65$ (a), $x=0.75$ (b), $x=0.80$ (c),
and $x=0.86$ (d). To compare the two nematic susceptibilities of
each composition, the upper panels show the temperature dependence
of $\chi_{{\rm nem}}-\chi_{0}$, with the Curie-Weiss fits (solid
lines). The lower panels depict the inverse susceptibilities, from
which the Curie-Weiss temperatures are extracted.}
\label{fig3} 
\end{figure}


The doping evolution of the elastoresistance is depicted in Fig.\,3(a)-(d).
For Rb concentrations up to $x=0.65$, $\chi_{{\rm nem}}^{[110]}$
remains considerably larger than $\chi_{{\rm nem}}^{[100]}$. For
$x=0.65$, the temperature dependence of $\chi_{{\rm nem}}^{[110]}$
continues to follow a Curie-Weiss behavior with a negative $T_{0}$,
indicating the dominance of $B_{2g}$ nematic fluctuations, but there
is no evidence for long-range nematic order {[}Fig.\,3(a){]}. As
the Rb concentration further increases ($x=0.75,0.80,$ and $0.86$),
the magnitude of $\chi_{{\rm nem}}^{[100]}$ gradually grows at the
expense of $\chi_{{\rm nem}}^{[110]}$, as shown in Fig.\,3(b)-(d),
reflecting the development of a $B_{1g}$ nematic instability near
$x=1$. Remarkably, for $x=0.80$ and $0.86$, the nematic susceptibilities
in the two channels have comparable magnitudes and similar temperature
dependencies at high temperatures. In this doping range, the system
displays long-range $B_{1g}$ nematic order, as signaled by the positive
$T_{0}$ and, more importantly, by the peak in $\chi_{{\rm nem}}^{[100]}(T)$,
which defines $T_{{\rm nem}}$. Interestingly, despite the onset of
long-range order in the $B_{1g}$ channel, $\chi_{{\rm nem}}^{[110]}$
continues to increase even below $T_{{\rm nem}}$, indicating the
importance of both nematic modes for all temperatures. In the $x$
range between $B_{1g}$ and $B_{2g}$ nematic orders, i.e., for $x=0.75$,
the absence of peaks in the nematic susceptibilities indicates the
absence of long-range nematic order. In addition, the two nematic
susceptibilities show almost identical temperature dependencies, with
very similar Curie-Weiss temperatures. The observed
comparable Curie-Weiss behavior in these two symmetry-irreducible
channels immediately indicates that the nematic fluctuations are isotropic,
in contrast to the strongly anisotropic fluctuations found for $x=0$
and $x=1$. Note that to establish that the nematic fluctuations are
nearly isotropic (i.e. XY-like), it is sufficient to show that the
susceptibilities in the two orthogonal channels, $B_{1g}$ and $B_{2g}$,
are very similar. This is because a director pointing along an arbitrary
in-plane direction can always be decomposed as a linear combination
of the $B_{1g}$ director and the $B_{2g}$ director. The situations
is analogous to the case of a nearly-XY ferromagnet, which displays
comparable susceptibilities $\chi_{\mathrm{FM}}^{xx}$ and $\chi_{\mathrm{FM}}^{yy}$.
Thus, our results provide direct evidence that the $x=0.75$ concentration
is very close to an XY-nematic state. 

\begin{figure}[t]
\begin{centering}
\includegraphics[width=1\linewidth]{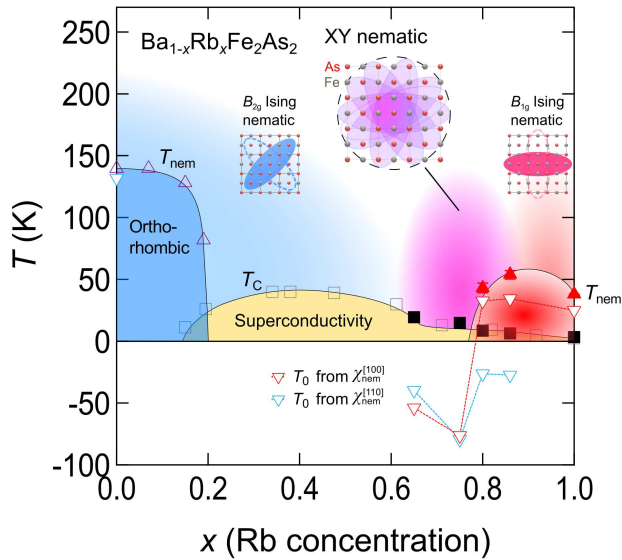} 
\par\end{centering}
\caption{ Temperature ($T$) versus Rb concentration ($x$) phase diagram obtained
in this study. The structural transition temperature $T_{{\rm nem}}$
(purple triangles) and the superconducting transition temperature
$T_{{\rm c}}$ (open squares) refer to previous results on polycrystalline
samples \cite{Peschke2014}. $T_{{\rm c}}$ of the single crystals
synthesized in this work are denoted by closed squares. The onset
temperature $T_{{\rm nem}}$ (red closed triangles) of the new $B_{1g}$
nematic state is defined by the peak in $\chi_{{\rm nem}}^{[100]}$.
The Curie-Weiss temperatures $T_{0}$ are extracted from the Curie-Weiss
fittings of $\chi_{{\rm nem}}^{[100]}$ (red inverted triangles) and
$\chi_{{\rm nem}}^{[110]}$ (blue inverted triangles). 
The lines are guides to the eyes. The blue, red, and purple shaded
regions highlight the three different types of nematic fluctuations
observed above the ordering temperatures. }
\label{fig4} 
\end{figure}


Our main findings are summarized in the Ba$_{1-x}$Rb$_{x}$Fe$_{2}$As$_{2}$
phase diagram shown in Fig.\,4. The electronic nematic director rotates
by $45^{\circ}$ with increasing Rb doping, changing from $B_{2g}$
Ising-nematic order near $x=0$ (blue region) to $B_{1g}$ Ising-nematic
order at $x=1$ (red region). In an intermediate doping range, the
nematic susceptibilities in the two channels become very similar,
thus revealing an emergent nearly-XY nematic fluctuating regime (purple
region). The Curie-Weiss temperature extracted from the $B_{1g}$
nematic susceptibility shows an abrupt change near the $x=0.80$ composition,
below which long-range $B_{1g}$ nematic order is no longer detected.
It is noteworthy that in the related compound Ba$_{1-x}$K$_{x}$Fe$_{2}$As$_{2}$,
the Fermi pockets near the Brillouin zone corner change from electron-like
to hole-like near a similar doping composition \cite{Xu2013}. 

Our results offer important new insights about the
microscopic mechanism of nematicity in iron-based superconductors.
More specifically, one can envision two possible general theoretical
scenarios to explain our results. In the first scenario, both the
$B_{2g}$ and $B_{1g}$ nematic states have the same origin. Given
the evidence in favor of a magnetically-driven $B_{2g}$ nematic state
for the $x=0$ compound \cite{Fernandes13}, the question is whether
magnetic fluctuations can also promote $B_{1g}$ nematicity. In this
regard, recent works have proposed that the wave-vector $\mathbf{Q}$
of the dominant magnetic fluctuations change from being parallel to
the Fe-Fe direction for $x=0$ to being parallel to the Fe-As direction
for $x=1$ \cite{Borisov2019,Wang19}. This would naturally lead to
a change in the vestigial nematic order from $B_{2g}$ to $B_{1g}$
\cite{Zhang17,Bishop17}. Indeed, first-principle calculations \cite{Borisov2019}
predict that the magnetic ground state changes from single-stripe,
$\mathbf{Q}=(\pi,\pi)$, for $x=0$, to double-stripe, $\mathbf{Q}=(\pi,0)$,
for $x=1$ (recall that we are referring to the crystallographic 2-Fe
unit cell). A related theoretical proposal also attributes the change
in nematicity to a change in the magnetic spectrum, which would become
incommensurate due to the occurrence of a Lifshitz transition as $x=1$
is approached \cite{Onari2018}. While long-range magnetic order has
not been observed in RbFe$_{2}$As$_{2}$, inelastic neutron scattering
measurements, which probe directly the magnetic fluctuation spectrum,
could be used to verify the applicability of this type of scenarios.

An alternative scenario is that the two nematic states
have different origins. In this context, recent studies of possible
charge order in RbFe$_{2}$As$_{2}$ \cite{Civardi2016} and in KFe$_{2}$As$_{2}$
under high pressure \cite{Wang2016} suggest that charge degrees of
freedom could be important. Moreover, a distinguishing feature of
the $x=1$ compound, and of all $3d^{5.5}$ stoichiometric compounds,
is the proposed enhanced strength of electronic correlations \cite{Hardy13,Eilers2016,Hardy2016,Mizukami2016,Medici2014,Valenti15}.
One interesting possibility then is that the $B_{1g}$ nematic state
in RbFe$_{2}$As$_{2}$ could be the result of a symmetry-breaking
Kondo-like hybridization gap between the more localized $d_{xy}$
orbital and the more itinerant $d_{xz}$, $d_{yz}$ orbitals \cite{Schmalian}.
Another possibility would be ferro-orbital order involving the $d_{xz}$,
$d_{yz}$ orbitals only, although a microscopic mechanism that does
not involve magnetic fluctuations has yet to be proposed.

\section{CONCLUSIONS}

From the elastoresistance measurements in Ba$_{1-x}$Rb$_{x}$Fe$_{2}$As$_{2}$, it is found that the nematic director rotates from Fe-Fe to Fe-As directions with hole doping. In the intermediate doping regimes, the nematic susceptibilities in the two symmetry-irreducible ($B_{1g}$ and $B_{2g}$) channels show very similar Curie-Weiss behavior, indicating nearly isotropic nematic fluctuations that are different from the Ising nematicity found in other iron-based superconductors.

The present observation of electronic XY-nematic fluctuations in overdoped
Ba$_{1-x}$Rb$_{x}$Fe$_{2}$As$_{2}$ offers a new route to realizing
this novel quantum liquid-crystalline state in iron-based superconductors.
A promising direction is attempting to stabilize long-range XY-nematic
order, for instance by combining chemical substitution and application
of pressure \cite{Borisov2019}. One of the unique properties of such
an XY-nematic state is that its Goldstone mode couples directly to
the electronic density, as discussed above. Of course, the system
will inevitably have a small anisotropy favoring one of the two nematic
components, which will gap the Goldstone mode. Yet, for a small anisotropy,
as implied by our elastoresistance data for the $x=0.75$ composition,
much of the system's behavior (except at very low energies) is similar
to that of a true XY-nematic state.

\section*{Acknowledgements}

We thank fruitful discussion with A.\,E. B\"{o}hmer, V. Borisov, A. Chubukov, A. Fujimori,
Y. Gallais, H. Kontani, C. Meingast, S. Onari, J.\,C. Palmstrom, I. Paul, J. Schmalian, Q. Si, 
 and R. Valenti. This work was supported by ``Kakehashi''
collaborative research program of Tsukuba Innovation Arena and by
Grants-in-Aid for Scientific Research (Nos.\
15H02106, 18H05227, 18K13492) from Japan Society for the Promotion
of Science. X-ray diffraction measurements were partly supported by
the joint research in the Institute for Solid State Physics, the University
of Tokyo. Theory work (R.M.F.) was supported by the U.S. Department
of Energy, Office of Science, Basic Energy Sciences, under Award No.\
DE-SC0012336.


\section*{Appendix A: Single crystals}

Single crystals of \RbAs \ and \CsAs \ were grown from arsenic-rich
flux as described in Ref.\,\onlinecite{Eilers2016}. The crystals
grown by this method have been characterized by several measurements
including X-ray, thermal expansion, ac susceptibility, and specific
heat \cite{Eilers2016,Hardy2016,Mizukami2016}, and quantum oscillations
have been observed in the magnetostriction \cite{Eilers2016} indicating
the cleanness of these samples. Single crystals of \BaRb \ with
various Rb concentrations $x$ were grown by the FeAs self-flux method
\cite{Tsujii}. 
$x$ was determined by energy dispersive X-ray spectrometry and also
checked from the $c$-axis lattice constant measured by X-ray diffraction.
The temperature dependence of the in-plane resistivity is measured
by the standard four-probe method, and the results of \BaRb \ single
crystals are shown in Fig.\,5. The superconducting transition temperature
\tc \ of the crystals was determined by the onset of superconducting
transition (black arrows). The characteristic S-shaped temperature
dependence as well as overall doping dependence of $T_{c}$ are similar
to that reported in \BaK \ single crystals \cite{Liu2014}.

\begin{figure}[t]
\includegraphics[width=0.8\linewidth]{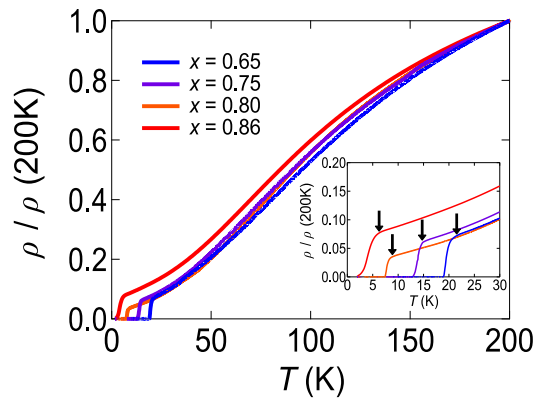} \caption{Temperature dependence of the resistivity for the \BaRb \ single
crystals. The vertical axis is normalized by the resistivity value
at 200\,K. The inset shows an expanded view below 30\,K. The arrows
indicate the onset of superconducting transition. }
\label{figS1} 
\end{figure}

\section*{Appendix B: Nematic susceptibility measurements}

\begin{figure}[b]
\includegraphics[width=1\linewidth]{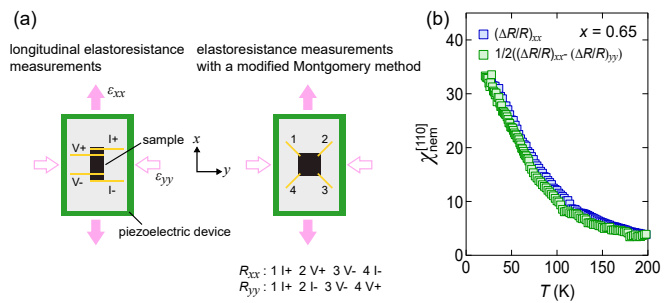} \caption{Biaxial strain effect to the nematic susceptibility. (a) Schematic
illustration of the two elastoresistance measurements. (b), The temperature
dependence of $\chi_{{\rm nem}}^{[110]}$ for Ba$_{1-x}$Rb$_{x}$Fe$_{2}$As$_{2}$
with $x=0.65$ obtained by longitudinal elastoresistance measurements
(blue square), which is the same data as in Fig.\,3(a), and by a
modified Montgomery method (green square).}
\label{figS2} 
\end{figure}

We have systematically evaluated the $B_{1g}$ and $B_{2g}$ nematic
susceptibility in $A$Fe$_{2}$As$_{2}$ ($A=$ Ba, Cs and Rb) and
Ba$_{1-x}$Rb$_{x}$Fe$_{2}$As$_{2}$ (Figs.\,2 and 3) from longitudinal
elastoresistance measurements, where the current direction is parallel
to the applied strain $\epsilon_{xx}$. At present, we have not obtained
reliable data for KFe$_{2}$As$_{2}$, whose crystals are more sensitive
to air.

For the longitudinal measurements we use rectangular crystals with
typical dimensions of $\sim1.0\times0.4\times0.02$\,mm$^{3}$. The
resistance of the sample ($R_{xx}$) was measured by the conventional
four-probe method {[}Fig.\,6(a){]}. For the $B_{1g}$ and $B_{2g}$
nematic susceptibility measurements, the edges of the samples are
aligned along the tetragonal {[}100{]} and {[}110{]} directions, respectively.

It should be noted that, strictly speaking, the strain transmitted
to the sample via the orthorhombic deformation of the piezoelectric
stack has not uniaxial but highly anisotropic biaxial character, which
can be decomposed into in-plane $A_{1g}$-symmetry strain ($\frac{1}{2}(\epsilon_{xx}+\epsilon_{yy})$)
and $B_{1g/2g}$-symmetry strain ($\frac{1}{2}(\epsilon_{xx}-\epsilon_{yy})$).
However, since $\epsilon_{xx}$ and $\epsilon_{yy}$ have opposite
signs, and their ratio is characterized by the in-plane Poisson's
ratio $\nu_{p}$ of the piezoelectric device ($\epsilon_{yy}=-\nu_{p}\epsilon_{xx}$),
the $A_{1g}$-symmetry strain is small compared to the $B_{1g/2g}$-symmetry
strain \cite{Kuo2013}. To see the effect of $A_{1g}$-symmetry strain
on the elastoresistance, we also conducted additional elastoresistance
measurements using a modified Montgomery method \cite{Kuo2016}. For
this type of measurement, samples with square shape were used ($\sim0.6\times0.6\times0.03$\,mm$^{3}$),
and the resistances along both $x$ and $y$ directions ($R_{xx}$
and $R_{yy}$) are measured by switching the current and voltage configurations
{[}Fig.\,6(a){]}. In this method, the nematic susceptibility $\chi_{{\rm nem}}$
is given by $\chi_{{\rm nem}}\sim(1/2((\Delta R/R)_{xx}-(\Delta R/R)_{yy}))/\epsilon_{xx}$,
in which $A_{1g}$-symmetry strain effect can be removed.

As shown in Fig.\,6(b), the data of $\chi_{{\rm nem}}^{[110]}(T)$
in Ba$_{1-x}$Rb$_{x}$Fe$_{2}$As$_{2}$ with $x=0.65$ determined
by two methods are almost identical, revealing that the $A_{1g}$-symmetry
strain effect is negligibly small in the present measurements, and
the longitudinal elastoresistance measurements can be used to extract
nematic susceptibility in our study.

\end{document}